\documentclass[11pt,a4paper]{article}
\usepackage{cite}
\usepackage{amsmath, amsthm, amssymb,slashed}
\usepackage{yfonts}

\usepackage{ifpdf}
\ifpdf
  \usepackage[pdftex]{graphicx}
  \usepackage{epstopdf}
\else
  \usepackage[dvips]{graphicx}
\fi

\usepackage{subfigure}
\usepackage{xfrac}  
\usepackage{fancyhdr}
\usepackage{mathbbol}
\usepackage{url}
\usepackage{color}
\usepackage{bbm}
\usepackage{rsfso}
\usepackage{dsfont}
\usepackage{bm}
\usepackage{amsfonts}
\usepackage{mathrsfs}
\usepackage{slashed}
\usepackage{wasysym}
\usepackage{yfonts}
\usepackage{ytableau}
\usepackage{tikz}
\usetikzlibrary{arrows,decorations.pathmorphing,hobby}
\usetikzlibrary{calc,arrows.meta}
\usepackage{cancel}
\usepackage[normalem]{ulem}

\def\bl{\begin{equation}\begin{aligned}}
\def\el{\end{aligned}\end{equation}}
\def\beal{\begin{align}}
\def\eal{\end{align}}
\def\be{\begin{equation}}
\def\ee{\end{equation}}
\def\bpm{\begin{pmatrix}}
\def\epm{\end{pmatrix}}
\def\bsm{\begin{bmatrix}}
\def\esm{\end{bmatrix}}
\def\bvm{\begin{vmatrix}}
\def\evm{\end{vmatrix}}
\def\bVM{\begin{Vmatrix}}
\def\eVM{\end{Vmatrix}}
\def\bea{\begin{eqnarray}}
\def\eea{\end{eqnarray}}

\def\1{{\bf 1}}
\def\2{{\bf 2}}
\def\3{{\bf 3}}
\def\4{{\bf 4}}

 \def\sut{SU(3)~}
  \def\sutf{SU(3)$_F$~}

\usepackage{indentfirst}

\newcommand{\OMIT}[1]{}

\usepackage{mathtools}
\usepackage{tikz}

\title{Open charm tetraquarks  in broken \sutf symmetry}
\date{\today}
\author{ L. Maiani$^{\dagger *}$, A.D. Polosa$^\dagger$ and V. Riquer$^{\dagger *}$ \\
  {\it $^\dagger$ Sapienza University of Rome and INFN, Piazzale Aldo Moro 2, I-00185, Italy.} \\ {\it $ {^*}$ CERN, 1211 Geneva 23, Switzerland. }}

\begin{document}
\maketitle

\begin{abstract}

Prompted by a recent lattice QCD calculation, 
we review the \sut light quark flavor structure of charmed tetraquarks with spin $0$ diquarks.
Fermi statistics forces the three light quarks to be in the representation $ {\bar {\bm{3}}}\otimes {\bar {\bm{3}}}= {\bm{3}} \oplus {\bar{\bm {6}}}$. This agrees  with the weak repulsion in the $\bm{15}$ of the $\bm {3}\otimes \bm{8}$ in $\bar DK$ scattering studied on the lattice. We analyze the ${\bm{3}} \oplus{ \bar{\bm  {6}}}$ multiplet broken by the strange quark mass and determine the five  independent masses from the known masses of diquarks. 
%
The mass of $D^* _{s0}(2317)$ 
is predicted within 50 MeV accuracy. The recently observed $\bar D_s^{--}(2900)$ and $\bar D_s^{0}(2900)$, likely part of a $I=1$ multiplet, with flavor composition $\bar c \bar q q^\prime s$, 
and $X_0(2900)$, an isosinglet with flavor composition $\bar c\bar s ud$, fit naturally in a ${\bm{3}} \oplus{ \bar{\bm  {6}}}$ structure as the first radial excitations. We discuss also the decay modes of $D^* _{s0}(2317)$, of the radial excitations and of the predicted particles.
\end{abstract}

\section{Introduction}

Charmed-strange tetraquarks are studied  in a recent lattice QCD calculation~\cite{Yeo:2024chk} in connection with the  \sutf configurations of possible bound states in the $\bar D K$ channel. Allowed \sutf multiplets are those appearing as irreducible components of the tensor product
\be
\bar DK= {\bm {3}}\otimes{\bm {8}}={\bm {3}}\oplus {\bar{\bm 6}}\oplus {\bm {15}} \label{bardk}
\ee
 Ref.~\cite{Yeo:2024chk} finds attraction in ${\bm {3}}$ and ${\bar{\bm 6}}$ but not in ${\bm {15}}$.
 
 Tetraquark of the same flavor have been considered earlier in connection with the SELEX observation of a charm-strange meson decaying into\footnote{We define: $\bar D^-_s=(\bar c s),\bar D=(\bar c q)$, $K=(\bar q s)$.}: $D^+_s+ \eta ~\text{or} ~D^0+K^+$~\cite{Maiani:2004xg}. 

With reference to \sutf, we consider here the antidiquark-diquark composition
\be
 [\bar c  \bar v]^{\bm{3}_c}_0~[  q_1 q_2]^{\bar{\bm 3}_c}_0 \label{tcs}
  \ee 
 where the subscript refers to spin zero and $(v, q_1,q_2=u,d,s) $.


 \section{Quantum numbers and states}
 
Fermi statistics requires the product $q_1q_2$ to be antisymmetric in flavour, it being already antisymmetric in spin (to get total spin 0) and color (to obtain a ${\bar{\bm{3}}_c}$). The corresponding \sutf multiplets are in the tensor product
\be
{\bar{\bm 3}}\otimes {\bar{\bm 3}}={\bm {3}}\oplus {\bar{\bm 6}} \label{tflav}
\ee
 the same attractive channels found in~\cite{Yeo:2024chk} and no $ {\bm {15}}$. 

Some authors have considered diquark-antidiquark states with diquarks in color ${\bm 6}$. Spin $0$ diquarks would be antisymmetric under spin $\times$ color exchange therefore they would be in a ${\bm {6}}$ representation of \sutf. Uncharmed, quarks  would belong then to the flavour representations ${\bar{\bm 3}}\otimes {\bm{6}}={\bm{3}}\oplus {\overline{\bm{15}}}$, in disagreement with~\cite{Yeo:2024chk}.

\begin{figure}[h]
\centering
\begin{tikzpicture}
    \draw[line width=1.5pt] (0,0) -- (4,0) -- (2,3) -- cycle;
    \draw[line width=1.5pt,dashed] (1,3/2) -- (3,3/2);
    \draw[line width=1.5pt,dashed] (1,3/2) -- (2,0);
     \draw[line width=1.5pt,dashed] (3,3/2) -- (2,0);
    \filldraw (0,0) circle (3pt);
     \filldraw (2,0) circle (3pt);
      \filldraw (4,0) circle (3pt);
       \filldraw (1,3/2) circle (3pt);
        \filldraw (3,3/2) circle (3pt);
         \filldraw (2,3) circle (3pt);
    \node at (0,-0.5) {$S_{11}$};
        \node at (2,-0.5) {$S_{12},T^3$};
            \node at (4,-0.5) {$S_{22}$};
    \node at (0.1,3/2) {$S_{13},T^2$}; 
     \node at (3.9,3/2) {$S_{23},T^1$}; 
       \node at (2,3.5) {$S_{33}$};         
        \draw[line width=1pt] (2,0) circle (0.17cm);
         \draw[line width=1pt] (1,3/2) circle (0.17cm);
          \draw[line width=1pt] (3,3/2) circle (0.17cm);
    \node at (6,0) {$S=-1$}; 
    \node at (6,3/2) {$S=0$}; 
     \node at (6,3) {$S=+1$};      
         \node at (-0.7,-1.5) {$I_3=-1$};
         \node at (0.8,-1.5) {$-1/2$}; 
    \node at (2,-1.5) {$0$};
    \node at (3,-1.5) {$1/2$}; 
     \node at (4,-1.5) {$+1$};      
\end{tikzpicture}
\caption{The $\bm 3 \oplus \bar{\bm 6}$ representation in  the $I_3$-Strangeness plane. Electric charges are as follows
$Q(S_{11})=-2$, $Q(S_{13})=Q(S_{12})=-1$ and  $Q(S_{33})=Q(S_{23})=Q(S_{22})=0$. \label{I3S}}
\end{figure}
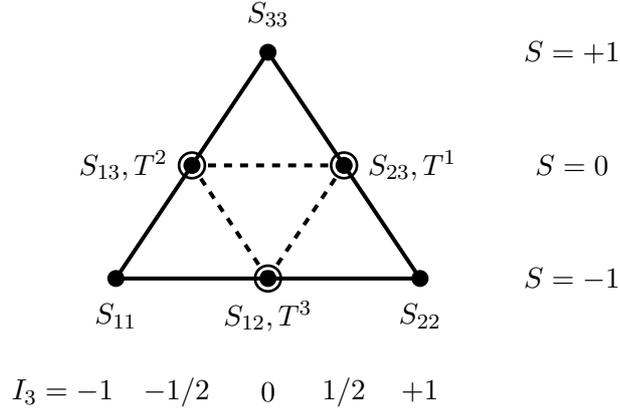

Let us find the explicit form of tetraquarks~\eqref{tcs}. We introduce the tensors $T^i$ in the $\bm 3_F$ representation and the tensors $S_{ij}$ in the $\bar{\bm 6}_F$ representation as~\footnote{The convention is that quarks (antiquarks) carry a upper (lower) flavor index.}:
\be
T^i=\bar v_\alpha (q^\beta q^\gamma) \epsilon_{\beta \gamma \delta}\epsilon^{\delta \alpha {\color{red} i}} \propto \bar v_\alpha q^\alpha q^i
\ee
since quark fields anticommute. The normalized vectors for triplet $(T)$ tetraquarks are (diquark spin $0$ understood)
\begin{align}
&S=0 & T^1=\frac{[\bar c \bar d][du]+[\bar c \bar s][su]}{\sqrt{2}} && T^2=\frac{[\bar c \bar u][ud]+[\bar c \bar s][sd]}{\sqrt{2}} \label{fthree0} \\
&&\notag \\
&S=-1& T^3=\frac{[\bar c \bar u][su]+[\bar c \bar d][sd]}{\sqrt{2}} \label{fthreem1} 
\end{align}
Similarly in the flavor sextet $(S)$ tetraquarks 
\be
S_{ij}=\frac{1}{2}[\bar v_i (q^\beta q^\gamma) \epsilon_{j \beta \gamma }+(i \leftrightarrow j)]
\label{sette}
\ee
and the normalized ${\bar {\bm{6}}}$ vectors are
\begin{align}
&S=+1 & S_{33}=[\bar c\bar s][ud] \label{six1}  \\
&S=0 & S_{13}=\frac{[\bar c\bar u][ud]+[\bar c\bar s][ds]}{\sqrt{2}} &&  S_{23}=\frac{[\bar c\bar d][ud]+[\bar c\bar s][su]}{\sqrt{2}}\label{six0norm} \\
&S=-1 & S_{11}=[\bar c\bar u][ds] && S_{12}=\frac{[\bar c\bar u][s u]-[\bar c\bar d][sd]}{\sqrt{2}} &&&& S_{22}=[\bar c\bar d][su] \label{sixm1norm}
\end{align}
In presence of \sutf breaking, $m_u=m_d<m_s $, we expect  the mass eigenstates with $S=0$ to correspond to the combinations
\be
S_{13}\pm T^2\qquad S_{23}\pm T^1 \label{mixed}
\ee
Fig.~\ref{I3S} gives the pattern of Sextet and Triplet  states in the $I_3$-Strangeness plane.


Following~\cite{Yeo:2024chk}, we identify  $T^3$ in Eq.~\eqref{fthreem1} with the observed $D^*_{s0}(2317)$~\cite{ParticleDataGroup:2022pth} (see also the review~\cite{nonqqbar}).
In Sect.~\ref{radial} we will discuss the particles observed by LHCb: $D_{s0}(2900)^0\to D^+_s \pi^-=[cd\bar s\bar u]$, $D_{s0}(2900)^{++}\to D^+_s \pi^+=[cu\bar s\bar d]$~\cite{LHCb:2022sfr}  and $X_0(2900)\to D^- K^+=[\bar c \bar s d u]$~\cite{LHCb:2020pxc}.

\section{Mass formulae in broken \sutf} \label{massf}

We introduce the symmetric masses with $M_{\bar{\bm{6}} }, M_{\bm{3}}$, and add octet \sutf  breaking using the symbols $m_{\bm 6}$ and $m_{\bm 3}$. 
In the product ${\bar{\bm{6}} }\otimes {\bm{6}}$ representation ${\bm{8}}$ appears only once, so there is only one operator to describe the symmetry breaking, namely the Hypercharge of the light quarks,  given by the formula
\be
Q_\ell=I_3 +\frac{1}{2}Y_\ell
\ee
and suffix $\ell$ means that we ignore the charm antiquark. For the  representation ${\bar{\bm{6}}}$
\be 
Y_{\ell,{\bar{\bm{6}}}}= \text{diag}\left(\frac{4}{3},\frac{1}{3},-\frac{2}{3}\right)~~\text{for}~~(S_{33},S_{12},S_{11})~~\text{and}~~\text{Tr}(Y_{\ell,{\bar{\bm{6}}}})=0
\ee
The symmetry breaking mass in the representation $\bar {\bm 6}$ is
\be
m_{\bar{\bm{6}}}=\beta_{\bar{\bm{6}}}\, \frac{1}{2}\left(Y_{\ell,{\bar{\bm{6}}}}+ \frac{2}{3}\right)
\ee
explicitely
\be
 m_{\bar{\bm{6}}}= \beta_{\bar{\bm{6}}}\, \text{diag}\left(1,\frac{1}{2},~0\right)~~\text{for}~~(S_{33},S_{12},S_{11}) \label{symbreak6}
 \ee
Similarly, for the ${\bm{3}}$ representation
\be
Y_{\ell,{{\bm{3}}}}=\text{diag} \left(\frac{1}{3},-\frac{2}{3}\right)~~\text{for}~~(T^{1},T^{3})~~\text{and}~~\text{Tr}(Y_{\ell,{\bar{\bm{3}}}})=0
\ee
with the symmetry breaking
\be
m_{\bm{3}}=\beta_{\bm{3}}\left(Y_{\ell,{{\bm{3}}}}+ \frac{2}{3}\right)=\beta_{\bm{3}}~\text{diag}(1,0)~~\text{for}~(T_1, T_3) \label{symbreak3}
\ee

Mixing ${\bm {{3}}}-{\bar{\bm {6}}}$  is described by the  by the matrix
\be
m_{\rm mix} \propto{ {\lambda}}_8 \label{mix}=\text{diag}(1,1,-2)
\ee
and the matrix ${\cal M}$ mixing $T^1,S_{23}$ or equivalently $T^2,S_{13}$ is
\be
{\cal M}=\bpm M_{\bm{3}} +\beta_{\bm {{3}}}& \delta\\ \delta &M_{\bar{\bm{6}}}+\frac{\beta_{\bar{\bm{6}}}}{2} \epm\label{mixm}
  \ee 

In total we have 5 states and 4 independent physical masses:  $i)~M(S_{33})$; $ii)$ and $iii)$ corresponding to the masses $M_{\pm}$ (see Eq.~\eqref{eval}) of the two $S=0$ states arising from the mixing matrix~\eqref{mixm}, and $iv)$ $M(S_{11})=M(T^3)$, since they have the same flavour composition. Enforcing the latter condition gives the relation
\be
M_{\bm {{3}}} =M_{\bar{\bm{6}}} \label{subst}
\ee
 and we remain with 4 parameters, $M_{\bar{\bm{6}}}=M, \beta_{\bm{3}},\beta_{\bar{\bm {6}}},\delta$.
The magic mixing in \eqref{mixed} is obtained for equal diagonal terms in Eq.~\eqref{mixm}, that is
\be
\beta_{\bm{3}}= \frac{\beta_{\bar{\bm{6}}}}{2}  \label{magmix}
\ee

To first order in $\beta_{\bm{3}}$ and $\beta_{\bar{\bm{6}}}$, eigenvalues and eigenstates of the mixing matrix~\eqref{mixed} with the substitution~\eqref{subst} are given by
\bea
&& M_\pm=M+\frac{2\beta_{\bm{3}}+\beta_{\bar{\bm{6}}}}{4} \pm \delta; \label{eval}
\eea

 In addition to the equality of $M(S_{11})$ and $M(T^3)$, the quark composition of the ${\bm{3}}\oplus{\bar{\bm{6}}}$ suggests an interesting regularity, namely  that $\beta_{\bm{3}}$ and $\beta_{\bar{\bm{6}}}$ have to be very small, if not vanishing at all.
Indeed, according to~\eqref{sette},  the lower indices in $S_{11}$ correspond to the quark-diquark antisymmetric configuration $\bar u \otimes [ds]_{\text{A}}$ while the  lower indices in $S_{33}$ correspond to $\bar s\otimes [ud]_{\text{A}}$ which have obviously the same content in quark masses, two light and one heavy.  

Exact equality of the bound states masses corresponds to $\beta_{\bm{3}}=\beta_{\bar{\bm{6}}}=0$: same masses at the upper vertex and lower corners of the triangle in Fig.~\ref{I3S}. 
In this case, symmetry breaking is restricted to the mass difference between the two $S=0,  I=1/2$ multiplets  induced by ${\bm {{3}}}-{\bar{\bm {6}}}$ mixing and of order $\mu\sim 2(m_s-m_q)$, with all other masses degenerate at $M$.

Small values of $\beta_{\bm{3}}~\text{and}~\beta_{\bar{\bm{6}}}$ could result from differences in the hyperfine interactions, which are between different pairs in the two cases (see below, Eq.~\eqref{massformula}).

The situation can be compared to the case of charmed baryons, where the two light quarks in spin one are also in a flavour symmetric ${\bm{6}}$ representation. In this case indices $1$ or $3$ univocally correspond to $u$ or $s$ quarks, and the top and bottom particles ($\Sigma_c$ and $\Omega_c$) differ in mass by $240$~MeV~\footnote{Baryon and meson spectroscopy suggests a value: $m_s-m_q\sim 170$~MeV (see e.g.~\cite{Ali:2019roi}), however the difference $M(\Omega_c)-M(\Sigma_c)$ receives a contribution of opposite sign from the hyperfine, spin-spin interaction.}, of the order of $2(m_s-m_q)$.

Group theory is effective at disentangling the ambiguity in these two cases by making use of the parameters allowed by the Wigner-Eckart theorem.

Another interesting case is that of hidden charm \sutf tetraquarks where a lower or upper index $3$ is unequivocally associated with a strange quark or antiquark and, correspondingly, the octet obeys the equal spacing rule of vector mesons, with spacing $\sim(m_s-m_q)$, well satisfied by the masses of $X(3872)-Z_{cs}(4003)-X(4140)$~\cite{Maiani:2021tri}.

\section{Comparing with the diquark-antidiquark model} \label{comparing}

Mass formulae for tetraquarks in terms of diquark masses and hyperfine interactions have been spelled in Ref.~\cite{Maiani:2014aja}, with reference to hidden charm tetraquarks.

For hyperfine interactions, the formula proposed in Ref.~\cite{Maiani:2014aja} is
\bea
 (H_{\rm h.f.})_{ij}&=&2\kappa_{ij}~({\bm s}_i\cdot {\bm { s}_j})=\kappa_{ij} \left[s (s+1)-\frac{3}{2}\right] \notag \\
\kappa_{ij}&=&\frac{|\Psi(0)|^2}{m_i m_j} \label{hhf}
  \eea
where $s$ is the total spin of the $ij$ pair belonging to the same diquark, under the assumption that the overlap probability for quarks in different diquarks is negligible.
This hypothesis reproduces the observed mass ordering: $X(3872),Z(3900)<Z(4020)$. 

\begin{table}[b]
\centering
    \begin{tabular}{||c|c|c|c||}
     \hline
quark & $q$& $s$ & $c$  \\ \hline
$q$ & $300\pm 100$ & $490\pm 10$ & $1877$  \\ \hline
 $s$ &$490\pm 10$ & $-$ & $2035$  \\ \hline
  $c$ &$1877$ & $2035$ & $-$  \\ \hline
\end{tabular}
 \caption{\footnotesize {Complete diquark masses, ${\overline{M}}_{ij}$,~in MeV.}}
\label{dqm}
\end{table}

To simplify the notation, we define ``complete diquark masses" which include the hyperfine interaction appropriate to diquarks with spin $=0$, e.g.
\be
{\overline M}_{cq} = M_{cq}- \frac{3}{2}\kappa_{cq} \text{,~etc.}
\ee
Numerical values are reported in Tab.~\ref{dqm}. Charmed diquark masses and hyperfine interactions are taken from Refs.~\cite{Maiani:2014aja,Maiani:2016wlq} and complete masses for uncharmed, spin $0$ diquarks from the, not so well determined, masses of the light scalar mesons~\cite{Jaffe:1976ih}, $f_0(500)$ and $f_0(980)$ (see the errors in Tab.~\ref{dqm})
\be
{\overline M}_{qq}=\frac{1}{2}M(f_0(500))\qquad {\overline M}_{sq}=\frac{1}{2}M(a_0(980)).
\ee

With reference to Eq.~\eqref{six1} and \eqref{sixm1norm} one has
\bea
&&M(S_{33})= {\overline M}_{cs}+{\overline M}_{qq}=M+\beta_{\bar{\bm{6}}} \\
&&M(S_{11})=M(T^3) ={\overline M}_{cq}+{\overline M}_{sq}=M
\eea
where we used the first and third entries respectively of $m_{\bar {\bm 6}}$ in  Eq.~\eqref{symbreak6}.
Here and in the following, we assume ${\overline M}_{cs}={\overline M}_{\bar c\bar s}$ etc. and $q=u,d$. Mixed states
\bea
&&M(S_{13})=M(S_{23})=M+\frac{1}{2}\beta_{\bar{\bm{6}}}\notag \\
&& M(T^1)=M(T^2)=M+\beta_{\bm{3}}
\eea
where we used the second entry of $m_{\bar{\bm 6}}$ in~\eqref{symbreak6} and the first entry of $m_{{\bm 3}}$ in~\eqref{symbreak3}. 
The sum of these two quantities is the trace of the matrix~\eqref{mixm}. 
Using~\eqref{mixed} and~\eqref{eval}
\be
  M_++M_-=2M+\frac{2\beta_{\bm{3}}+\beta_{\bar{\bm{6}}}}{2}
  \ee
that is
\be
\frac{2\beta_{\bm{3}}+\beta_{\bar{\bm{6}}}}{2}=M_++M_- -2M=[({\overline M}_{cs}-{\overline M}_{cq})-({\overline M}_{sq}-{\overline M}_{qq})]
\ee
given that (from~\eqref{six0norm}) $M_{+}+M_-=\overline M_{cs}+\overline M_{sq}+\overline M_{cq}+\overline M_{qq}$. 
 Similarly:
\be
M(S_{33})-M(T^3)=\beta_{\bar{\bm{6}}}=[({\overline M}_{cs}-{\overline M}_{cq})-({\overline M}_{sq}-{\overline M}_{qq})] 
\label{betap}
\ee
and we find
\be
 \beta_{\bar{\bm{6}}}=-32 \pm 100~\text{MeV}
 \ee
Predicted masses of ${\bm {3}}$ and ${\bar{\bm{6}}}$ are (use values in Table~1)
\bea
&&M(S_{11})=M(T_3)=\overline M_{cu}+ \overline M_{sd}=2367\pm 10~\text{MeV}\notag \\
&& M(T_-)=\overline M_{cu}+ \overline M_{ud}=2177 \pm 100~\text{MeV}\notag \\
 && M(T_+)=\overline M_{cs}+ \overline M_{sd}=2525 \pm 10~\text{MeV}\notag \\
 &&M(S_{33})=\overline M_{cs}+  \overline M_{ud}=2335\pm 100~\text{MeV}\label{massformula}
\eea
The first value compares favorably with the mass of the observed $D^{*-}_{s 0}(2317)$, with a difference of $50\pm 10$ MeV. 

\section{The multiplet of radial excitations} \label{radial}

The particles $D^0_{s0}(2900)\to D^+_s \pi^-=[cd\bar s\bar u]$, $D^{++}_{s0}(2900)\to D^+_s \pi^+=[cu\bar s\bar d]$ recently observed by LHCb~\cite{LHCb:2022sfr} and $X_0(2900)\to D^- K^+=[\bar c \bar s d u],~$,~\cite{LHCb:2020pxc} are too heavy to be included in the basic ${\bm{3}}\oplus {\bar{\bm{6}}}$ together with $D^*_{s0}(2317)$. The mass difference:
\be
M(2900)-M(2317)=583~\text{MeV}
\ee
is similar to the mass gap between $\psi(2S)$ and $J/\psi$ ($\Delta=590$~MeV) or between $X(3872)$ and $Z(4430)$ ($\Delta=558$~MeV) and we shall similarly interpret the LHCb resonances as the first radial excitations $(n=2)$ of the basic multiplet  the $D^*_{s 0}(2317)$ belongs to.

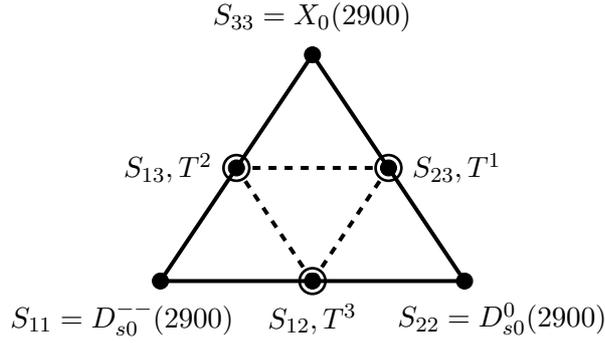
\begin{figure}[htb!]
\centering
\begin{tikzpicture}
    \draw[line width=1.5pt] (0,0) -- (4,0) -- (2,3) -- cycle;
    \draw[line width=1.5pt,dashed] (1,3/2) -- (3,3/2);
    \draw[line width=1.5pt,dashed] (1,3/2) -- (2,0);
     \draw[line width=1.5pt,dashed] (3,3/2) -- (2,0);
    \filldraw (0,0) circle (3pt);
     \filldraw (2,0) circle (3pt);
      \filldraw (4,0) circle (3pt);
       \filldraw (1,3/2) circle (3pt);
        \filldraw (3,3/2) circle (3pt);
         \filldraw (2,3) circle (3pt);
    \node at (-0.5,-0.5) {$S_{11}=D^{--}_{s0}(2900)$};
        \node at (2,-0.5) {$S_{12},T^3$};
            \node at (4.5,-0.5) {$S_{22}=D^{0}_{s0}(2900)$};
    \node at (0.1,3/2) {$S_{13},T^2$}; 
     \node at (3.9,3/2) {$S_{23},T^1$}; 
       \node at (2,3.5) {$S_{33}=X_0(2900)$};         
        \draw[line width=1pt] (2,0) circle (0.17cm);
         \draw[line width=1pt] (1,3/2) circle (0.17cm);
          \draw[line width=1pt] (3,3/2) circle (0.17cm);
\end{tikzpicture}
\caption{The $n=2$ multiplet. $\bar D_s\pi, ~S=-1$ and $\bar D K,~S=+1$ resonances observed by LHCb~\cite{LHCb:2022sfr,LHCb:2020pxc} in the $n=2$ multiplet. \label{I3Sn2}}
\end{figure}

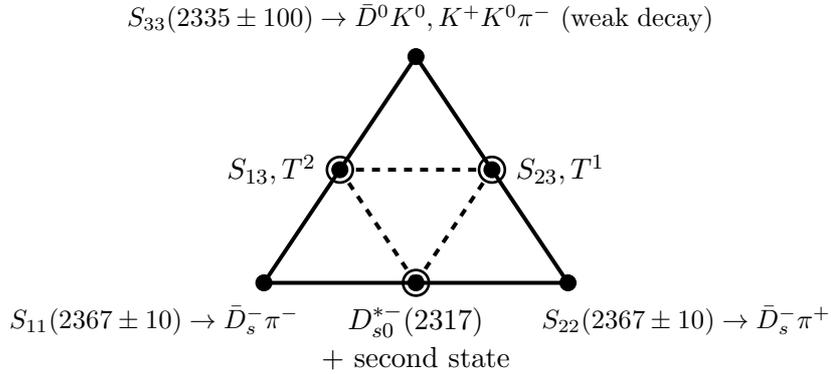
\begin{figure}[htb!]
\centering
\begin{tikzpicture}
    \draw[line width=1.5pt] (0,0) -- (4,0) -- (2,3) -- cycle;
    \draw[line width=1.5pt,dashed] (1,3/2) -- (3,3/2);
    \draw[line width=1.5pt,dashed] (1,3/2) -- (2,0);
     \draw[line width=1.5pt,dashed] (3,3/2) -- (2,0);
    \filldraw (0,0) circle (3pt);
     \filldraw (2,0) circle (3pt);
      \filldraw (4,0) circle (3pt);
       \filldraw (1,3/2) circle (3pt);
        \filldraw (3,3/2) circle (3pt);
         \filldraw (2,3) circle (3pt);
    \node at (-1.5,-0.5){\small{ {$S_{11}(2367\pm10)\to \bar D_s^-\pi^-$}}};
        \node at (2,-0.5) {$D_{s0}^{*-}(2317)$};
         \node at (2,-1) {$+$ second state};
            \node at (5.5,-0.5){\small{ {$S_{22}(2367\pm 10)\to \bar D_s^-\pi^+$}}};
    \node at (0.1,3/2) {$S_{13},T^2$}; 
     \node at (3.9,3/2) {$S_{23},T^1$}; 
       \node at (2,3.5){\small{ {$S_{33}(2335\pm 100)\to \bar D^0 K^0, K^+K^0\pi^- $}~(weak decay)}};         
        \draw[line width=1pt] (2,0) circle (0.17cm);
         \draw[line width=1pt] (1,3/2) circle (0.17cm);
          \draw[line width=1pt] (3,3/2) circle (0.17cm);
\end{tikzpicture}
\caption{The $n=1$ multiplet. The diquarks in $S_{23}$ are $[\bar c\bar s][su](2525\pm 10)\to \bar D_s^-K^0, \bar D^0\eta$ and $[\bar c\bar d][ud](2177\pm 100)\to \bar D^0\pi^0$\label{missing}}
\end{figure}

We have to fit in the same multiplet $X_0(2900)$ with $D^{--,0}_{s0}(2900)$, antiparticles of the resonances observed in~\cite{LHCb:2022sfr}, in order to have the same charm quantum number,  see Fig.~\ref{I3Sn2}.
The expected $n=1$ multiplet is shown in Fig.~\ref{missing} 

The positive strangeness $X_0(2900)$ mass close to the masses of the negative strangeness particle $D^{--,0}_{s0}(2900)$ is a remarkable confirmation of the regularity noted in  Section~\ref{massf}, a real footprint of the tetraquark compositions: $[\bar c\bar s]_0[ud]_0$ and $[\bar c\bar u]_0[sd]_0$.

\section{Decays}

{\bf \emph{{ The case of $D^-_{s 0}(2317)$}.}} 
As shown by Eq.~\eqref{fthreem1}, $T^3$  has $I=0$ and it should decay into $D_s^- \eta$, which however is forbidden by phase space.  We can consider two independent mechanisms for the observed, isospin violating, $D_s^- \pi^0$ decay, both related to the $m_d-m_u$ mass difference: mixing of $T^{3}$ with $S_{12}$ ($I=1,I_3=0 $), or $\eta-\pi^0$ mixing.

 In both cases, mixing allows the decay $D^*_{s0}\to D_s \pi^0$ with a small width ($\Gamma<3.8$~MeV is reported in~\cite{ParticleDataGroup:2022pth}).  
It would be interesting to observe the decay $D^*_{s0}\to D_s \gamma\gamma$, quoted in ~\cite{ParticleDataGroup:2022pth} with an upper bound to the branching ratio $B(\gamma\gamma)<0.18$, to compare with $D^*_{s0}(2317) \to D^-_s \eta^* \to  D^-_s  \gamma\gamma$ via the virtual $\eta$.

\vskip01cm
{\bf \emph{{The missing partners of ${\bm{D^-_{s 0}(2317).}}$}}} With reference  to Fig.~\ref{missing}, the bottom corners must be filled by two, isovector mesons in the channels $D^-_s\pi^\pm$, in all similar to those found at mass $2900$~MeV in~\cite{LHCb:2022sfr}. In addition a companion of $D^*_{s0}(2317)$ is needed, close in mass and in the same channels: $\bar D^-_s \pi^0 ~\text{or} ~\gamma \gamma$, most likely with a larger width.

The lighter, zero strangeness state, predicted at $2177$, could be identified with the lower pole under $D^*(2300)$ reported in PDG~\cite{ParticleDataGroup:2022pth} at mass $2105$.

The most intriguing case is the particle in the upper vertex, which is predicted to be very close to the $\bar D K$ threshold, the channel where  $X_0(2900)$ is seen. If it is below the threshold of this channel, it has to decay weakly into $K^+ K^0\pi^-$.

{\bf\emph{{Radial excitations.}}} With the larger mass of the radial excitations shown in Fig.~\ref{I3Sn2} all possible two body decays are open:
\bea
&& (S_{12},~T^3)_{(n=2)} \to D_s^- \pi^0,~D_s^- \eta, \notag \\
&& (S_{12},~T^3)_{(n=2) }\to \bar D^0 K^-, \bar D^- \bar K^0
\eea

The mixing of $n=2$ states $S_{12}$ and $T^3$ can be determined from the decay rates as in Ref.~\cite{Maiani:2004xg}. 

For zero strangeness states, we expect the OZI mixing to produce tetraquarks with and without one $s\bar s$ pair:
\bea
&&[\bar c \bar s][s d]_{(n=2)}\to  \bar D^- \eta,~\bar D^-_s K^0,  \notag \\
&& [\bar c \bar u][ u d]_{(n=2)},~ [\bar c \bar d][ u d]_{(n=2)} \to \bar D \pi.
\eea
\section{The role of Fermi statistics in single charm tetraquarks}

In a very interesting paper, the authors of Ref.~\cite{Zhang:2024fxy} utilize the so-called light quark spin symmetry in the static quark approximation~\cite{Isgur:1989vq}
 to classify spin states of hidden charm molecules of quark composition $(\bar c q)(\bar q^\prime c)$, with fixed Isospin $I$. Calling $S_{\ell,I}$ and $S_{c\bar c} (=1,0)$ the light quarks and $c\bar c$ total spin, the possible combinations of light and heavy spin generate six states with definite Isospin, total angular momentum and charge conjugation:  $J^{PC}_I= 0^{++}_I, 1^{+-}_I,1^{\prime+-}_I, 1^{++}, 0^{\prime ++}_I, 2^{++}_I$. No surprise, these are the same six $J^{PC}_I$ states produced by diquark-antidiquark  color singlet tetraquarks of the form $[cq]^{\bar{\bm {3}}} [\bar c{\bar q}^\prime]^{\bm {3}}$, considered in~\cite{Maiani:2004xg,Maiani:2014aja}. 
 
 The situation is different in the case considered in Eq.~\eqref{tcs} of the present paper. 
 Assuminig diquark $\otimes$ antidiquark colors to be ${\bar{\bm{3}}}\otimes {\bm {3}} \to {\bm{1}}$, there is a correlation between total spin and  Isospin (or $SU(3)$ flavour) of the light quarks pair $q_1 q_2$ induced by Fermi statistics. The latter requires either (a): ${\bar{\bm{3}}}_f \leftrightarrow (S_{12}=0)$ or (b): ${\bm{6}}_f \leftrightarrow (S_{12}=1)$. Therefore, in the case at hand we are led univocally to flavour  ${\bar{\bm{3}}}_f$ and to the ${{\bm{3}}}_f\oplus {\bar{\bm{6}}}_f$ composition of the tetraquark structure studied in this paper.
 
 The situation is different  for the molecular structure $(\bar c q_1)(\bar v q_2)$, in that the colors of $q_1$ and $q_2$ are not correlated and there are no apparent reasons for spin $0$ molecules  not to display all flavours in the representations appearing in the $SU(3)$ flavour decomposition of the product $\bar D K$,~Eq.~\eqref{bardk}.
 
For $J^P=0^+$ single charm exotics, the suppression of the ${\bf{15}}$ was derived in Ref.~\cite{alabadejo} for the molecular case, from an explicit calculation using chiral dynamics, following the lines indicated in~\cite{Kolomeitsev:2003ac}.
 \section{Conclusions}
We show how the resonance $D_{s0}^{--,0}(2900)$ and $X_0(2900)$ nicely fit in a $\bar{\bm 6}$  representation of \sutf with the prediction of a few more states in the sextet, in addition to the very likely  $D_{s0}^{-}(2900)$ to fill an isotriplet with $D_{s0}^{--,0}(2900)$. The observation that $M(2900)-M(2317)=583~\text{MeV}\simeq M(\psi(2S))-M(\psi(1S))$ suggests that the sextet we discuss could be a radial excitation of a lower sextet containing the $D_{s0}^*(2317)$, in a similar way in which $Z(4430)$ can be interpreted as a radial excitation of $X(3872)$\cite{Maiani:2014aja}.  Using \sutf symmetry breaking we obtain mass predictions for the missing states. Our results are in agreement with a recent lattice calculation\cite{Yeo:2024chk} showing that in the $\bar D K$ scattering there are no bound states in the $\bm{15}$ representation, something which is expected in the quark model description we present here. 

\section*{Acknowledgements}
We are grateful to Vanya Belyaev and Tim Gershon for a very useful discussion of LHCb results on the single charm, single strange mesons discussed here. We are indebted to  Christofer Hanhart for pointing us the lattice result in~\cite{Yeo:2024chk} that has been at the origin of the present article and for an interesting discussion on the molecular picture of exotic hadrons.


\begin{thebibliography}{99}
\bibitem{Yeo:2024chk}
J.~D.~E.~Yeo, C.~E.~Thomas and D.~J.~Wilson,
[arXiv:2403.10498 [hep-lat]].
\bibitem{Maiani:2004xg}
L.~Maiani, F.~Piccinini, A.~D.~Polosa and V.~Riquer,
Phys. Rev. D \textbf{70} (2004), 054009
doi:10.1103/PhysRevD.70.054009
[arXiv:hep-ph/0407025 [hep-ph]].
\bibitem{Maiani:2014aja}
L.~Maiani, F.~Piccinini, A.~D.~Polosa and V.~Riquer,
Phys. Rev. D \textbf{89} (2014), 114010

[arXiv:1405.1551 [hep-ph]].
\bibitem{Maiani:2016wlq}
L.~Maiani, A.~D.~Polosa and V.~Riquer,
Phys. Rev. D \textbf{94} (2016) , 054026

[arXiv:1607.02405 [hep-ph]].
 
 \bibitem{Jaffe:1976ih}
R.~L.~Jaffe,
Phys. Rev. D \textbf{15} (1977), 281
doi:10.1103/PhysRevD.15.281

%
\bibitem{ParticleDataGroup:2022pth}
R.~L.~Workman \textit{et al.} [Particle Data Group],

doi:10.1093/ptep/ptac097

PTEP \textbf{2022} (2022), 083C01

\bibitem{LHCb:2022sfr}
R.~Aaij \textit{et al.} [LHCb],
Phys. Rev. Lett. \textbf{131} (2023) 041902;

\bibitem{LHCb:2020pxc}
R.~Aaij \textit{et al.} [LHCb],
Phys. Rev. D \textbf{102} (2020), 112003

[arXiv:2009.00026 [hep-ex]].


 \bibitem{Ali:2019roi}
A.~Ali, L.~Maiani and A.~D.~Polosa,
Cambridge University Press, 2019,

ISBN 978-1-316-76146-5, 978-1-107-17158-9, 978-1-316-77419-9

doi:10.1017/9781316761465.

\bibitem{Maiani:2021tri}
L.~Maiani, A.~D.~Polosa and V.~Riquer,
Sci. Bull. \textbf{66} (2021), 1616

[arXiv:2103.08331 [hep-ph]].




\bibitem{nonqqbar} S. Eidelman {\it et al}., {\it Heavy Non-$q\bar q$ Mesons},  RPP 2022.



arXiv:2212.02716 [hep-ex].

\bibitem{Maiani:2004uc}
L.~Maiani, F.~Piccinini, A.~D.~Polosa and V.~Riquer,
Phys. Rev. Lett. \textbf{93} (2004), 212002

[arXiv:hep-ph/0407017 [hep-ph]].

\bibitem{Zhang:2024fxy}
Z.~H.~Zhang, T.~Ji, X.~K.~Dong, F.~K.~Guo, C.~Hanhart, U.~G.~Mei\ss{}ner and A.~Rusetsky,
[arXiv:2404.11215 [hep-ph]].

\bibitem{Isgur:1989vq}
N.~Isgur and M.~B.~Wise,
Phys. Lett. B \textbf{232} (1989), 113-117
doi:10.1016/0370-2693(89)90566-2


\bibitem{alabadejo} M.~Albaladejo, P.~Fernandez-Soler, F.~K.~Guo and J.~Nieves,
Phys. Lett. B \textbf{767} (2017), 465-469
doi:10.1016/j.physletb.2017.02.036
[arXiv:1610.06727 [hep-ph]].

\bibitem{Kolomeitsev:2003ac}
E.~E.~Kolomeitsev and M.~F.~M.~Lutz,
Phys. Lett. B \textbf{582} (2004), 39-48
doi:10.1016/j.physletb.2003.10.118
[arXiv:hep-ph/0307133 [hep-ph]].


\end{thebibliography}
 \end{document}